# Using fluctuations of Entropy as a starting point for obtaining behavior of scalar fields permitting collapse of thin wall approximation

A. W. Beckwith

Department of Physics and Texas Center for Superconductivity and Advanced Materials at the University of Houston
Houston, Texas 77204-5005 USA

**ABSTRACT**

We have shown that a scalar field can be used, employing Scherrer's k essence cosmological sound calculation, to model how we evolve from a dark matter-dark energy mix to a cosmological constant. Here, we are exploring what initiates the decay of the near perfect thin wall approximation on that scalar field to circumstances permitting a k essence speed-of-sound argument in favor of traditional models of Einstein's cosmological constant as a driving force for inflationary expansion.

Correspondence: A. W. Beckwith:    projectbeckwith2@yahoo.com

# INTRODUCTION

We have described [1] how a modification of the thin wall model for the scalar field approximation will lead to the k-essence speed-of-sound calculation changing from Scherrer's [2] conclusions about a dark energy-dark matter mix to a cosmological constant treatment of expansion. Next, we explore what initiates the decay of the near perfect thin wall approximation on that scalar field. Such circumstances permit a k essence speed-of-sound argument in favor of a cosmological constant. We identify initial fluctuations in entropy as a significant factor in changing the k essence speed of sound to one permitting more traditional contributions of a scalar potential that approaches Guth's chaotic inflationary cosmology model. This assumes, as well, that this variation of scalar potential would be accompanied by declining scalar field values, thereby leading to an effective constant scalar potential.

# OBTAINING FLUCTUATIONS OF ENTROPY DUE TO VARIATIONS OF PHASE

In an article about holography and pre big bang state of matter, A.K. Biswas and J. Maharana [3] derive a linkage between Planck's minimum lenth $l_P$, enthropy $S$, and a so called area A value, which in zeroth order is

$$l_P^2(t) \equiv \left(\frac{S}{A}\right)^{-1} \frac{1}{\sqrt{3}\beta T} \tag{1}$$

with $\frac{1}{\sqrt{3}\beta T}$ a constant. The time dependence of $l_P$ is an artifact of string theory. They specifically state that the ratio of $\left(\frac{S}{A}\right)$ is a constant due to $\frac{1}{\sqrt{3}\beta T}$.

I use their linkage to obtain fluctuations of entropy due to variations of phase but differ from Biswas and Maharna[3] by an infinitesimal amount of variation of the entropy S.

Note also that

$$l_P \equiv \hbar \cdot G \tag{2}$$

I am assuming, for this pre big bang situation, the Brans-Dicke variation of the gravitational factor, G via

$$G \propto \frac{1}{\langle \phi \rangle} \tag{3}$$

where

$$\langle \phi \rangle \equiv \langle \phi_0 \rangle + \langle \tilde{\phi} \rangle \tag{4}$$

with

$$\langle \phi_0 \rangle \gg \langle \tilde{\phi} \rangle \tag{5}$$

In this approach, $\langle \tilde{\phi} \rangle$ is linked to the scalar field used for the prior write up[1] about a background scalar field. This $\langle \tilde{\phi} \rangle$ decays from a near perfect thin wall approximation to a characterization that permits approaching the speed of sound contribution to k essence physics, which permits the Einstein cosmological constant to be a major factor in inflationary cosmology.[1,2]

We can, in this situation, make the following analogy,

$$\frac{\hbar^2}{\langle \phi_0 \rangle^2} \cdot \left(1 - 2 \cdot \frac{\langle \tilde{\phi} \rangle}{\langle \phi_0 \rangle}\right) \approx l_P^2(t) \equiv \left(\frac{S_0}{A}\right)^{-1} \frac{(1 - \delta \cdot S/S_0)}{\sqrt{3}\beta T} \tag{6}$$

$$\Rightarrow \frac{\delta \cdot S}{S_0} \sim \frac{\langle \tilde{\phi} \rangle}{\langle \phi_0 \rangle} \tag{7}$$

So, then we have

$$\langle \tilde{\phi} \rangle \cong (\delta \cdot t) \cdot \langle \dot{\phi} \rangle \sim \delta \cdot S \tag{8}$$

and, for the scalar field, use a quasi-one-dimensional respesentation[1,2]

$$\phi \sim N(t) \cdot \pi \cdot [\tanh(b(t) \cdot (x + L(t)/2)) - \tanh(b(t) \cdot (x - L(t)/2))] \tag{9}$$

where $N$ gets smaller with increasing time, while $b$ and $L$ get larger.

The point is that we obtain a situation where the fluctuation in entropy is the same as when we have fluctuations in the observed behavior of the scalar field. This is done in order to present relevant cosmological issues when representing a physical scalar field right after the nucleation of a new universe.

## CONCLUSION

In prior analyses,[1,4] we examined an alteration of an effective potential along the lines of

$$V \approx V_1 \cdot (1 - \cos \phi) + V_2 \cdot (\phi - \phi^*)^2 \xrightarrow[v \to l \arg er]{} V_3 \left( \frac{m^2 \phi^2}{1 + V_4 \cdot m^3 \cdot \phi^3} \right) \sim m^2 \cdot \phi^2 \tag{10}$$

This assumed that $V_1 \gg V_2$ and that the initial potential corresponded to a rising nucleation value for the scalar potential. Whereas after that nucleation phase occurred, we assumed that we would have a steady decay of the scalar potential after we went to the phase transition indicated by Eq. (10). At this step, the false vacuum nucleation procedure has been finalized for creation of the scalar field but there is a phase change to a declining overall magnitude of the height of the scalar potential, which is in tandem with a topological charge serving as a de facto screening effect. As this topological charge breaks down, Eq. (10) then implies an approach to traditional chaotic inflationary expansion. This is in tandem with the breakdown of the scalar potential from a thin wall approximation to one with a far sharper slope phase values, as seen for Eq. (9), as *b(t)* grows larger, and *N(t)* declines in magnitude.

The final step of the thin wall approximation breaking down has[1]

$$m^2 \cdot \phi^2 \xrightarrow[\phi \to \varepsilon^+]{} V_0 \equiv \text{constant} \tag{11}$$

Our analysis is that an increase in entropy is a causative basis for the thin wall approximation to break down. Furthermore, we are also asserting, in tandem with $G \propto \frac{1}{\langle \phi \rangle}$, that initial phases of entropy fluctuation also imply, at least in the beginning of the phase change represented in Eq. (10), that we are seeing a slight fluctuation in the gravitational constant values before the potential system of this problem decays into the conditions for implying the cosmological constant condition.[1]

Biswas[3] makes much of Eq. (1) being modified by string initial conditions. Our analysis states that one does not need cosmic strings per se, but that we have a major change in state of scalar field contributions to inflationary cosmology that deserve to be more fully investigated.

Note also that this minimum length, $l_P$, can be referred to as a minimum fluctuation length. This point is made by John Baez[5] who argued for

$$\Delta l \geq l_P \tag{12}$$

as a minimum fluctuation length occurring in the measurement of distances in cosmology. Putting together our prior work with nucleation of a scalar field shows that we are also investigating transforming from a vacuum state[1,4] to an initial condition presaging inflationary cosmology.[1,4] The minimum-distance uncertainty given by Eq. (12)[5] represents the order of magnitude of separation of constituent components of the scalar field. We are moving toward more traditional inflationary cosmology models due to the onset of fluctuations of entropy in our initial thin wall approximation. This has ramifications which will be explored in future research.

# REFERENCES

[1] arXIV: math-ph/0412002,' Matching Scherrer's k essence argument with behavior of scalar fields permitting derivation of a cosmological constant.' AW.Beckwith.

[2] arXIV astro-ph/0402316 v3 , May 6, 2004: Purely kinetic k-essence as unified dark matter R.J. Scherrer.

[3] arXIV :hep-th/99100106 v1 14 Oct 1999, The holography hypothesis in pre big bang cosmology with string sources. A K Biswas, J. Maarana.

[4] arXIV math-ph/0410060; How false vacuum synthesis of a universe sets initial conditions which permit the onset of variations of a nucleation rate per Hubble volume per Hubble time. A.W. Beckwith.

[5] arXIV:gr-qc/0201030 v1 , Jan 9, 2003. Uncertainty in measurement of distance. J. Baez, S.J. Olson.